\documentclass[a4paper,aps,preprint]{revtex4}
\usepackage{graphicx}
\usepackage{pstricks}
\usepackage{epsfig}
\usepackage{amsmath}
\begin{document}
\title{Non-Invasive Glucose Monitoring Techniques:\\ A review and current trends}

\author{Raju Poddar}
\altaffiliation{On Study Leave from Department of Biotechnology, 
Birla Inst. Tech., Mesra, Ranchi 835 215 India.}

\affiliation{Photonics Lab I, Division of Microelectorics, School of EEE, Nayang Technological University, Singapore - 639798.}

\author{Joseph Thomas Andrews}
\email[Corresponding author: ]{ jtandrews@sgsits.ac.in, Phone: +91-731-243 4095, www.sgsits.ac.in}
\affiliation{Department of Applied Physics,\\ Shri G S Institute of Technology
\& Science, \\ Indore - 452 003, India. } 

\author{Pratyoosh Shukla}
\affiliation{Department of Biotechnology, 
Birla Institute of Technology, Mesra, Ranchi 835 215 India.}

\author{Pratima Sen}
\affiliation{Laser Bhawan, School of Physics, Devi Ahilya University, \\ Khandwa Road, Indore 452 007 India.}

\begin{abstract}
Diabetes mellitus is a complex group of syndromes that have in common a disturbance in the body's use of glucose, resulting in an elevated blood sugar. Once detected, sugar diabetes can be controlled by an appropriate regimen that should include diet therapy, a weight reduction program for those persons who are overweight, a program of exercise and insulin injections or oral drugs to lower blood glucose. Blood glucose monitoring by the patient and the physician is an important aspect in the control of the devastating complications (heart disease, blindness, kidney failure or amputations) due to the disease. 
Intensive therapy and frequent glucose testing has numerous benefits. With ever improving advances in diagnostic technology, the race for the next generation of bloodless, painless, accurate glucose instruments has begun. In this paper,
we reviewed various methods, techniques and approaches successfully demonstrated
for measuring or monitoring blood glucose.
Invasive, minimally invasive and noninvasive techniques available in literature
are summarised. \end{abstract}

\maketitle
\section{Introduction}

Diabetes mellitus is a medical condition in which the body does not adequately produce the quantity or quality of insulin needed to maintain a normal circulating blood glucose. Insulin is a hormone that enables glucose (sugar) to enter the body's cells to be used for energy. Two types of diabetes are common. Type - I is also known as Insulin Dependent Diabetes Mellitus (IDDM) and accounts for 5-10\%  of all cases. Type - II or Non-Insulin Dependent Diabetes Mellitus (NIDDM) occurs in 90-95\% of the diabetic population. IDDM occurs in childhood. It requires insulin doses to maintain life, in addition to healthy practices. 
Frequent self-monitoring of blood glucose is crucial for effective treatment and reduction of the morbidity and mortality of diabetes. 

Unmonitored diabetes can lead to severe complications over time, including blindness, kidney failure, heart failure, and peripheral neuropathy associated with limb pain, poor circulation, gangrene and subsequent amputation \cite {Davidson}. These complications are largely due to poor glucose control. The Diabetes Care and Complications Trial (DCCT) demonstrate that more frequent monitoring of blood glucose and insulin levels could prevent many of the long term complications of diabetes \cite{Auxter}. However, the conventional blood (finger stick) glucose testing \& monitoring are painful, inconvenient due to disruption of daily life. Also, it causes fear of hyoglycemia resulting from tighter glucose control and may be difficult to perform in long term diabetic patients due to calluses on the fingers and poor circulation. 
A glucose measurement with following qualities, i) non-invasive, ii) non-contact, iii) fast measurement capability, iv) painless measurements, v) convenience for glucose monitoring and vi) cost effective which could provide adequate control and greatly reduce the complications seen in these patients. 

At present, the simplest and less painful method for glucose measurements are done by pricking a finger and extracting a drop of blood ($50\mu l/dl$) which is applied to a test strip composed of chemicals sensitive to the glucose in the blood sample. An optical meter (glucometer) is used to analyze the blood sample and gives a numerical glucose reading.

This paper refers the widely used monitoring techniques of blood glucose monitoring. A summary of the techniques discussed are given as a chart in figure 1.

\begin{figure}[htb]
\includegraphics[width=6in]{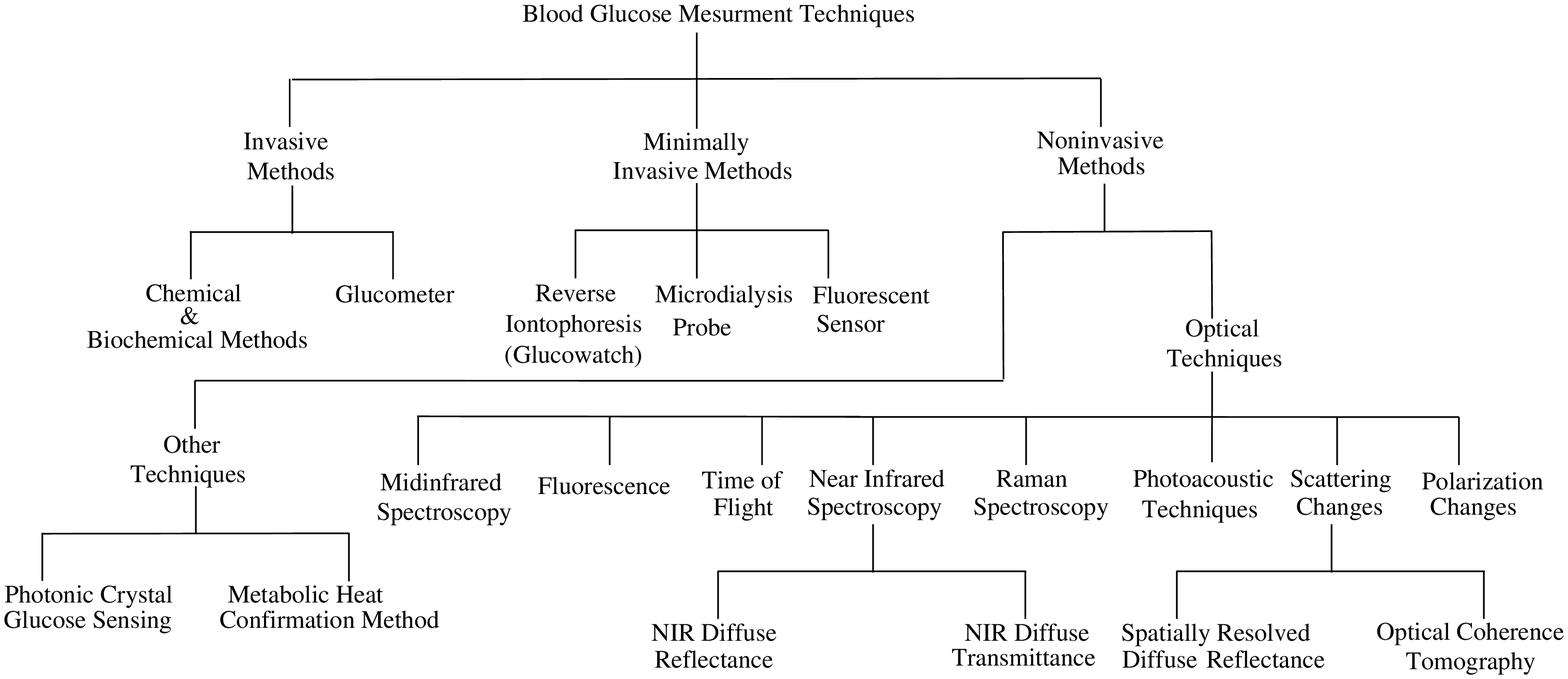}
\caption{\sf Methods of glucose measurement techniques.} 
\label{oct1}
\end{figure}

\subsection{Glucose in human body}
D-glucose is a molecule with the chemical formula C$_{6}$H$_{12}$O$_{6}$. in human body, food is converted into sugar and provides energy to all tissues and organs through blood circulation. In terms of its chemical composition, human blood sugar consists of D-glucose that exists mainly in the water base of blood plasma. The daily variation of glucose concentration in the human body is in the range of 60 - 160 mg/dl \cite{Auses}. Arterial and capillary blood taken from the fingertip have an identical glucose content, while the glucose level of venous blood is lower than the corresponding arterial value (1 - 17 mg/dl in healthy subjects and up to 30 mg/dl in diabetic patients). Besides, blood glucose also exists in other biofluids such as intracellular fluid, interstitial fluid, humour, saliva, sweet and urine. Researchers have established that, in the steady state condition, the glucose level in the intracellular and interstitial fluid is identical with the concentration of glucose in the blood. It is also known that the glucose level in humour correlates strongly with the glucose content of blood, while the glucose level in saliva, sweet and urine does not. It is the basic energy source for cellular metabolism. Glucose concentration varies between different anatomic regions and in different parts of the blood circulation. 
        
\begin{figure}[ht]
\includegraphics[width=5in,angle=0]{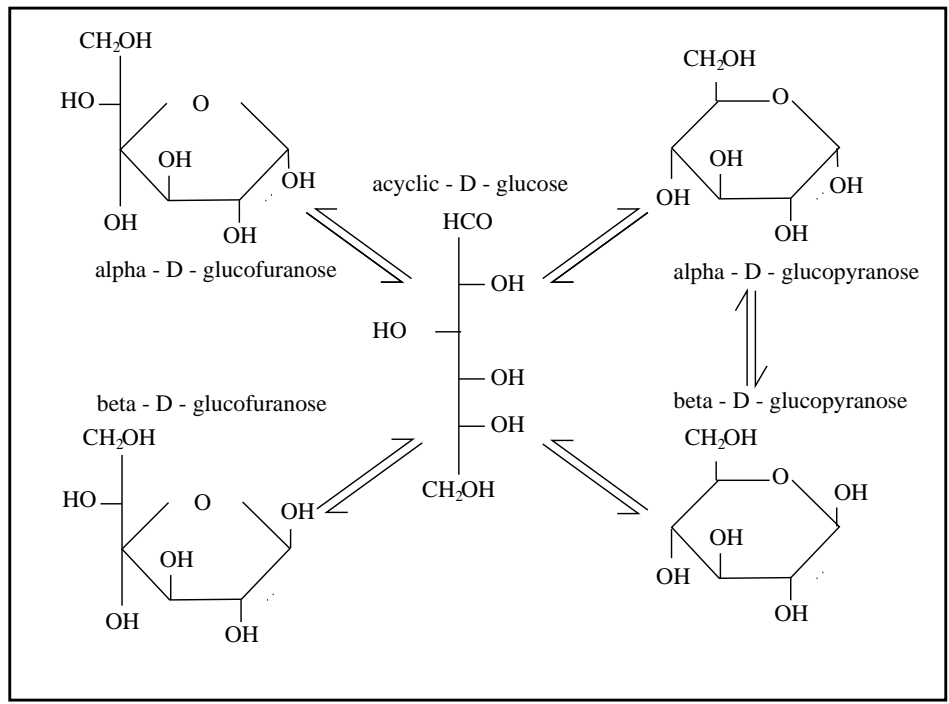}
\caption{\sf Anomers of D-glucose in an aqueous solution} 
\label{oct2}
\end{figure}

Glucose permeates red blood cells (RBCs) via passive diffusion, which is supported by 55-kD-glycoprotein \cite{Eggins}. RBCs work as a buffer in blood to control the plasma glucose concentration \cite{Shults}. It is suspected that D-glucose does not need a transporter to move between blood plasma and interstitial fluid (ISF). This process is thought to be driven by diffusion gradients. When the glucose concentration of plasma increases, the initial response is water movement from ISF to plasma; then the glucose diffuses into the ISF, where it is used as an energy source by cells. Because of these gradients, the glucose concentration in tissues is not a constant. The glucose concentration is higher in arterial blood than in venous blood, and the venous glucose concentration depends on the arterial blood flow rate and the rate of glucose uptake by the tissue \cite{Eggins}. Unlike between blood plasma and ISF, there are many glucose transporters through the cell membrane. The transport process of glucose is called passive-mediated or facilitated diffusion. In many organelles these transporters are stimulated by insulin. The distribution of the transporters varies between organelles. The transport rate of glucose into cells is possibly limited by glucose transporters or by glucose phosphorylation \cite{Shults}. In muscle cells, the glucose is phosphorylated rapidly after the intake into the cell \cite{Shults}.  Although many body fluids and tissues have been studied for non-invasive glucose sensing, in the best case scenario, the blood glucose concentration would be measured directly from the blood vessels. To make a diagnosis of Diabetes Mellitus, the vein plasma glucose should be measured \cite{Fischer}.

\subsection {Glucose oxidation methods}
Traditional glucose determination is based on the glucose oxidation reaction, catalyzed by glucose oxidase (GOD). It may be described by

\begin{equation}
\hspace{-5mm}\hbox{glucose} + O_{2} + H_{2}O \;\; \underrightarrow{\;\;\;\;GOD\;\;\;\;}\;\;\;  
H_{2}O_{2}
+ \hbox{gluconic  acid} 
\end{equation}

Glucose oxidase catalyzes the oxidation of $\beta$-D-glucose to D-gluconic acid and hydrogen peroxide. It is highly specific for $\beta$-D-glucose and does not act on $\alpha$-D-glucose. Its  major use is in the determination of free glucose in body fluids. Although specific for $\beta$-D-glucose, glucose oxidase can be used to measure the total amount of glucose. This is because, following the consumption of $\beta$-glucose, $\alpha$-glucose at equilibrium is converted to the $\beta$-form by mutarotation. The consumed oxygen or the ensuing production of gluconic acid or hydrogen peroxide (H$_{2}$O$_{2}$) is in direct proportion to the glucose content. The glucose oxidase method is characterized by high sensitivity, accuracy and reliability. Traditional electrochemical methods such as potentiometry or amperometry can be used to determine the glucose content during the glucose oxidation reaction. 

Very recently, the results obtained on the glycemia measurement by an indirect
method are presented.\cite{amar} The measurement method uses a sensor to
electromagnetic coupling based on loss currents. This sensor allows the detection of the blood's glucose level through the variation of the sensor impedance that is depend on the dielectric parameters
of blood, in particular the conductivity. The in vitro results presented a standard deviation of 2.1 mV, an average error
of 1.8 mV, and a maximum variation of 3.6 mV equivalents to a maximum error of $\pm$ 197 mg dl$^{-1}$ on
the concentration of D$^+$ glucose compared to the line of tendency.
 
\section{Non-Invasive Optical Techniques}

\subsection {General Features of Non-Invasive Optical Techniques}
Non-invasive optical measurement of glucose is performed by focusing a beam of light onto the body. The light is modified by the tissue after transmission through the target area. An optical fingerprint of the tissue content is produced by the diffuse light that escapes the tissue it has penetrated. The absorbance of light by the skin is due to its chemical components (i.e., water, hemoglobin, melanin, fat and glucose). The transmission of light at each wavelength is a function of thickness, color and structure of the skin, bone, blood and other material through which the light passes \cite{Jagemann}. 
The glucose concentration can be determined by analyzing the optical signal changes in wavelength, polarization or intensity of light. The sample volume measured by these methods depends on the measurement site. The correlation with blood glucose is based on the percent of fluid sample that is interstitial, intracellular or capillary blood.  The fluid viewed through the limb is 63\% intracellular and 37\% extracellular, of which 27\% is interstitial and 10\% plasma. A blood glucose value of 100mg/dl is equivalent to a tissue sample glucose average of 38 mg/dl of which 26\% is due to blood, 58\% is due to interstitial fluid and 16\% is due to intracellular fluid. 
Optical measurements dependent on concentration changes in all body compartments measured, as well as changes in the ratio of tissue fluids (as altered by activity level, diet or hormone fluctuations) and this, in turn, effects the glucose measurement. Problems also occur due to changes in the tissue after the original calibration and the lack of transferability of calibration from one part of the body to another. Tissue changes include: body fluid source of the blood supply for the body fluid being measured, medications that affect the ratio of tissue fluids, day-to-day changes in the vasculature, the aging process, diseases and the person's metabolic activity. However, the ratio of body fluids (intracellular, interstitial, plasma) are affected by factors such as activity level, diet or hormone fluctuations, but also by blood circulation, body temperature shift, metabolic activity and medication. All these factors are capable of influencing the optical parameters and, consequently, impacting the blood glucose measurement. Moreover, day-to-day changes in vasculature and tissue texture as well as the aging process may affect the long-term stability of glucose monitoring.

\subsection{Types of Measurement Techniques} 
Non-invasive glucose monitoring techniques can be grouped as subcutaneous, dermal, epidermal and combined dermal and epidermal glucose measurements. As all methods based on glucose oxidase require a direct contact between glucose and
some chemical reagents, they necessitate the extraction of glucose from the body. However, a non-contact, non-invasive method is impossible with any chemical based method. The only attraction is use a spectroscopic methods. In a spectroscopic technique is used an optical beam interacts with glucose within the human body. The generated signals are then analyzed and the results displayed. Since optical methods do not require the extraction of glucose from the body, they are highly suitable for the continuous, non-invasive monitoring of glucose.

Matrices other than blood under investigation include interstitial fluid, ocular fluids and sweat. Test sites being explored include finger tips, cuticle, finger web, forearm and ear lobe. Subcutaneous measurements include microdialysis, wick extraction, and implanted electrochemical or competitive fluorescence sensors. Microdialysis is also an investigational dermal and epidermal glucose measurement technique. Epidermal measurements can be obtained via infrared spectroscopy, as well. Combined dermal and epidermal fluid glucose measurements include extraction fluid techniques (iontophoresis, skin suction and suction effusion techniques) and optical techniques. Different properties of light are measured and the effect of glucose on the collected signals is evaluated with different techniques. Bruulsema et. al \cite{Bruulsema1} have studied the effect of glucose and the corresponding changes in optical properties. Glucose affects light scattering and also, to some extent, light absorption. The most pronounced effect of increased glucose concentration is decreased scattering coefficients. This overview is focused on a description of the optical techniques currently under development by diagnostic equipment manufacturers for glucose monitoring in diabetics - the fastest growing segment of diagnostic testing.
Till today the noninvasive blood glucose measurements techniques may categorized into two major groups viz. fluid glucose optical measurement strategies, fluid glucose biochemical measurement strategies and combination of both strategies. A summary of the optical techniques and their salient feature are summarize in the Table 1

Further, fluid glucose optical measurement techniques may be classified into the following subclasses,
\begin{enumerate}
        \item {Near Infrared Spectroscopy (NIR)}
        \item {Infrared Spectroscopy (IR)}
        \item {Raman Spectroscopy}
        \item {Photoacoustic and optoacoustic techniques}
        \item {Spatially resolved diffuse reflectance measurements}
        \item {Frequency-domain reflectance technique}
        \item {Polarization Changes}
        \item {Fluorescence}
        \item {Time of Flight (TOF) measurements}
        \item {Optical Coherence Tomography}
        \end{enumerate}

And, fluid glucose biochemical measurement strategies are,
        \begin{enumerate}
        \item {Photonic Crystal Glucose-Sensing Material}
        \item {Reverse iontophoresis followed by Glucose oxidase enzyme treatment used in Glucowatch}
        \item {Implantation of catheters in the area of subcutaneous layer of skin and reaction with Glucose oxidase enzyme used in Glucometer}
        \item {Measurement of Glucose by Metabolic Heat Conformation Method}
        \end{enumerate}

\section{Fluid Glucose Optical Measurement Strategies }

\subsection{Near Infrared Spectroscopy (NIR) }
Glucose produces one of the weakest NIR absorption signals per concentration unit of the body's major components. NIR spectroscopy measurement enables investigation of tissue depths in the range of 1 to 100 millimeters with a general decrease in penetration depth as the wavelength value is increased. NIR transmission through an ear lobe, finger web and finger cuticle or reflected from the skin of the forearm and lip mucosa has been attempted in the NIR region of $ 1\mu m - 2\mu m$. NIR diffuse reflectance measurements performed on the finger and cuticle have shown good correlation with blood glucose but 10\% of the predictions are not clinically acceptable \cite{Jagemann}.

NIR spectroscopy is based on collecting reflectance or absorption spectra of the tissue with a spectrometer. When near-infrared light illuminates a spot on the skin, the light is partially absorbed and scattered as a result of its interaction with chemical components in the illuminated tissue. Light that is not absorbed will be reflected out of the tissue or transmitted through it, before being received by optical detectors. An analysis of changes in the intensity of the light combined with the application of multivariate calibration techniques \cite{Haaland} permits the extraction of the tissue's chemical components, including glucose. Robinson \cite{Robinson} used NIR transmission spectroscopy and the multivariate calibration method to measure blood/tissue glucose concentrations in diabetic subjects. Several conventional spectrometer configurations are designed on transmission sampling in the 600 - 1300 nm range. The reported average prediction errors vary from 19.8 mg/dl to 37.8 mg/dl in oral glucose tolerance tests (OGTT) conducted on three subjects. In turn, Burmeister and Arnold \cite{Burmeister},  employed near-infrared transmission spectroscopy to study noninvasive blood glucose sensing in different measurement sites like cheek, lip, nasal septum, tongue and webbing tissue. The results showed that the tongue, containing the least amount of fat, provided spectra with the highest signal-to-noise ratio. They collected the transmission spectra of the tongue in the 1400 - 2000 nm ranges. Their findings showed a standard error of prediction (SEP) in excess of 54 mg/dl for all diabetic subjects \cite{Blank}. Hiese and Marbach \cite{Heise} have reported a series of studies on glucose determination in the oral mucosa membrane in the 1111 - 1835 nm spectral range through a diffuse reflectance measurement. The best SEP was 43 mg/dl from a 2-day single-person OGTT. Jagemann and cowokers \cite{Fischbacher}used a fiber optic probe to study diffuse reflectance over the 800 - 1350 nm range on the middle finger of the right hand. The blood glucose concentrations of the test persons were perturbed using carbohydrate loading. The results were evaluated using the partial least squares regression method (PLS) and radial basis function neural networks. In these tests, the mean root square prediction error was 3.6 mg/dl.This method has the advantage that no reagents are required for the measurements and that fiber optical components can be used. As a consequence only insulators are in direct contact with the skin. Furthermore the spectrometer can be constructed without moving parts, leading to a rugged design. 

\textbf{Methodology} Continuous wave light source is used to illuminate the tissue under study. A typical wavelength range 1050 - 2450 nm is used for measurements. The short wavelength region of the near infrared ($\lambda \cong$ 800-1300 nm) has the following advantages that are ideal for non-invasive and \textit{in vivo} diagnostics. In this range there is an optical window between the region in which visible light is absorbed by blood and skin pigments and the longer wavelengths in which absorption of water predominates. The NIR spectrum consists of overtones and combinations of the fundamental vibrations mainly of the bonds of carbon, nitrogen and oxygen with hydrogen. NIR spectra of aqueous systems show weak, broad and overlapping bands with random baselines. The position and intensity of the signals vary according to the chemical vicinity (hydrogen bonding effects). The influence of dissolved salts and temperature on the NIR spectra of aqueous systems is well known. Since the normal proportion of glucose in blood and tissue is only about 0.1 percent of the water content the spectral variations due to glucose concentration are extremely small. The evaluation of the recorded spectra is further complicated by several influences: water is a strong absorber and also the main component of living tissue. In addition time dependent biological processes take place, e.g. pulse and respiration. Other sources of variation are erroneous spectral recordings caused for example by irregular pressure of the measuring head on the finger. As a consequence multivariate calibration methods have to be used for evaluating the spectra.
\begin{figure}[ht]
\centering{\includegraphics[height=5in,angle=90]{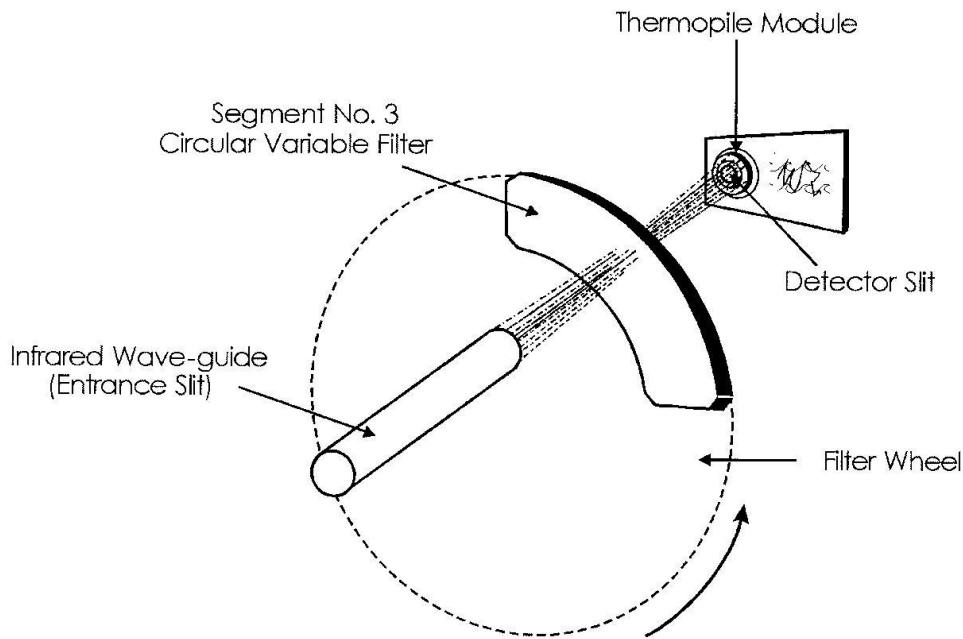}}
\caption{\sf Schematic of the filter-based IR spectroradiometer.} 
\label{oct3}
\end{figure}

\textbf{NIR spectrometer}: The equipment for recording the spectra is shown in Fig.  and consists of a light source, a fiber optical measuring head and a NIR spectrometer. This spectrometer uses a polychromator with a holographic imaging diffraction grating and an InGaAs photo diode array detector with 128 pixels mounted on a glass block. A very compact design (70 mm x 50 mm x 40 mm) with excellent optical performance was achieved by using a high-dispersion grating with extremely short focal length. The optical resolution (Rayleigh criterion) is about 12 nm in the wavelength region from 820-1320 nm. Fiber optical bundles connect light source (tungsten halogen lamp) and spectrometer with the fiber optical measuring head. The fibers illuminating the skin (finger tip) are concentrically arranged around the central part of the bundle which connects to the spectrometer. 
They compute the calibration coefficients by recording of the spectra from several diabetes patients with varying BG concentration levels.  The BG level is determined simultaneously by a conventional glucose analyzer using blood plasma. A typical blood glucose profile is shown in Fig. . The matrices of the calibration coefficients are computed by partial least-squares regression (PLS) and radial-basis neural networks (RBF).
\begin{figure}[ht]
\centering{\includegraphics[width=5in]{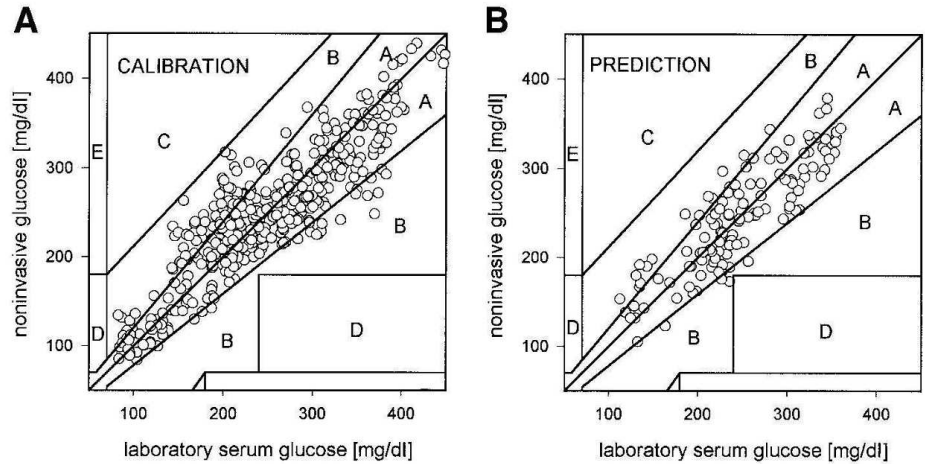}}
\caption{\sf Plot of correlations between invasive (laboratory) and noninvasive glucose concentrations for calibration (A) and prediction (B).} 
\label{oct4}
\end{figure}

Diffuse reflectance studies of the inner lip also have shown good correlation with blood glucose and indicated a time lag of 10 minutes between blood glucose and the measurement signal \cite{Marbach}. Salivary glucose levels (a component of lip measurements) did not reflect blood glucose levels. Physical and chemical parameters such as variation in pressure, temperature, triglyceride and albumin interfere with glucose measurement. Errors can also occur due to environmental variations such as changes in temperature, humidity, skin hydration, carbon dioxide, and atmospheric pressure. Extensive validation and testing of the glucose prediction equation is needed to determine if the glucose correlation is consistent in all clinically important conditions in all types of patients. dation and testing of the glucose prediction equation is needed to determine if the glucose correlation is consistent in all clinically important conditions in all types of patients. 

\subsection{Infrared Spectroscopy (IR)}
The IR glucose measurement systems at the epidermal surface enables investigation of tissue depths in the range of 10 to 50 micrometers at using a wavelength band in the IR region from 700 to 1000nm \cite{Kajiwara}. These systems are not suitable for measuring glucose in the blood containing tissues. An attenuated total reflection technique has been used for oral mucosa.

\textbf{Spectral Regions}: The entire near infrared region of the electromagnetic spectrum encompasses light with wavelengths ranging from 0.7 to 2.5 microns (14,286 - 4000 $cm^{-1}$ wavenumbers). For the most part, near infrared spectroscopic information corresponds to harmonics of overtones and combinations of fundamental vibrational transitions more frequently associated with mid-infrared spectroscopy. Overtone and combination absorptions are principally seen for CH, OH, and NH molecular groups. The energetics associated with these transitions result in absorption bands that are broad and featureless with low absorptivities. The ambiguity of CH, OH, and NH groups within biological systems and the physical nature of these transitions result in complex, overlapping spectra. The chemical environment surrounding these CH, OH, and NH groups controls the exact position and shape of these near infrared absorption features, thereby resulting in unique spectral signatures for the molecular species of interest. Selective analytical measurements rely on the uniqueness of these spectra. 
IR spectroscopy uses wavelengths between 2.5 – 20 mm. It is based on studies of light absorption by glucose at a selected wavelength range. The absorption spectrum are collected with a spectrophotometer. Zeller et al \cite{Zeller,Mendelson,Klonoff} have measured blood glucose concentration with IR spectroscopy.

Water is a critical matrix component for near infrared spectra of aqueous based clinical samples, such as the human body. The high concentration of water in clinical samples coupled with the relatively strong absorptivity of OH groups result in large water absorbance bands. The strong absorbance of water dictates using the regions between these water bands where sufficient amounts of light are transmitted. The following three regions are generally accessible: (i) the combination region: 2.0 - 2.5 microns (5000 - 4000 $cm^{-1}$); (ii) the first overtone region: 1.54 - 1.82 microns (6500 - 5500 $cm^{-1}$); and (iii) the short-wavelength near infrared (sw-NIR) region: 0.7 - 1.33 microns (14,286 - 7500 $cm^{-1}$). Glucose has three absorption bands in both the combination region (centered at 2.10, 2.27, and 2.32 microns) and the first overtone region (centered at 1.73, 1.69, and 1.61 microns). Although glucose absorption bands are difficult to measure in the sw-NIR owing to their extremely low absorptivities, bands centered at 0.76, 0.92, and 1.00 microns are reported.

A regression method is used to identify the best combination of tissue thickness to accurately simulate human \textit{in vivo} spectra. In this method, absorbance spectra from pure samples of the aqueous buffer and beef fat are combined according to equation : 
\begin {equation}
A_{H} = \beta_{0} + \beta_{1} * A_{w} + \beta_{2}*A_{f}
\end {equation}

where $A_{H},\; A_{w},$ and $A_{f}$ correspond to absorbance spectra for these pure samples of human webbing, water, and beef fat, respectively, and $\beta_{i}$ values correspond to the respective regression coefficients. Absorbance values ($A$) are defined according to equation : 
        \begin {equation}
A = - \log(I/I_{o})
\end {equation}

where $I$ and $I_{o}$ represent transmitted light intensities with and without the sample of interest, respectively. Absorbances are used owing to their additive nature according to the Beer-Lambert relationship. Application of this method involves starting with an \textit{in vivo} absorbance spectrum from the subject of interest and then fitting this spectrum by adjusting the relative amounts of the pure water and fat absorbance spectra in such a way to minimize the sum of the square of residuals. Model layer thicknesses are computed as the product of the regression coefficient and the known thickness of the corresponding pure samples. The regression model presented here represents a slightly simplified version where absorbance from muscle protein is ignored. This simplification is well justified for noninvasive spectra over the overtone region. Muscle protein absorbance must be considered, however, in the combination region. Quantitatively, we can compute the standard error of prediction (SEP) for the validation data in both cases. The SEP drops from 1.60 mM to 0.49 mM when increasing the path length from 5.6 to 6.2 mm. It must be noted that the spectral noise is essentially identical for these two data sets (23.4 and 23.5 micro-absorbance units for the 5.6 and 6.2 mm path length data, respectively). Improvements are expected from the Beer-Lambert relationship which states that longer path lengths will provide larger signals, thereby enhancing the signal-to-noise ratio of the measurement and improving analytical performance. 

\textbf{Experimental Parameters:} It is important to recognize that noninvasive blood glucose measurements are simply absorbance measurements in a complex matrix. As such, one must be able to differentiate the amount of light absorbed by glucose from spectral noise. In this regard sample thickness is a critical experimental parameter because tissue thickness affects both glucose sensitivity and spectral noise. As the sample thickness increases more light is absorbed for a given concentration of glucose, thereby enhancing sensitivity and lowering the detection limit. On the other hand, fewer photons successfully traverse a thicker layer of tissue, thereby reducing the measured light intensity and increasing spectral noise. A compromise is required to maximize sensitivity to glucose while minimizing spectral noise.
however, the drawbacks include glucose contamination of the measurement site by food and a highly variable saliva of low rate \cite{Hall}. Assays using whole blood as the sample matrix, are subject to interferences due to albumin, red cells and gamma globulin and changes in temperature and pH. Further, saliva glucose varies considerably and does not reflect blood glucose methods.

\begin{figure}[ht]
\includegraphics[width=5in]{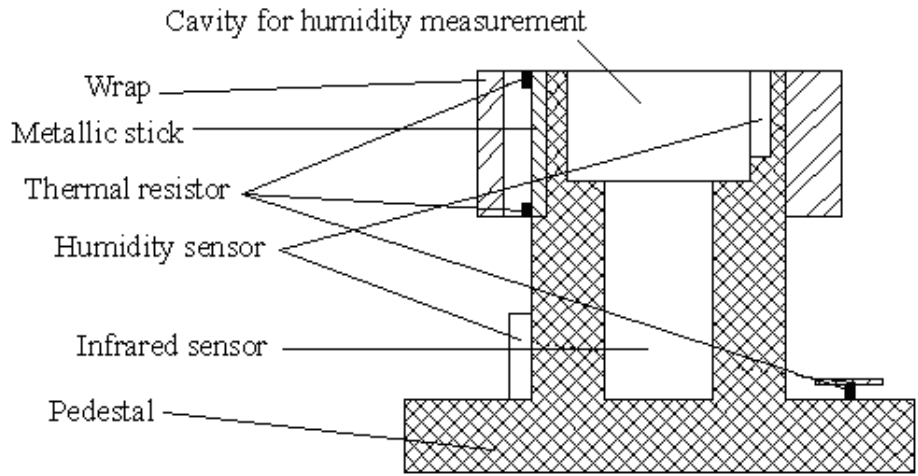}
\caption{\sf Structure of the detecting head for measurement blood glucose
by use of thermal diffusion.} \label{mhc}
\end{figure}

Very recently, a non-invasive glucose measurement system based on the method of metabolic heat conformation (MHC) is presented \cite{mhc}.
It consists (Figure \ref{mhc}) of three
temperature sensors, two humidity sensors, an infrared sensor and an optical measurement device. The glucose level can be deduced from the quantity of heat dissipation, blood flow rate of local tissue and degree of blood oxygen saturation. The correlation coefficient between the blood flow rates by this method and the results of a Doppler blood flow meter was found be equal to 0.914.

\subsection{Raman Spectroscopy}
The phenomenon of Raman scattering is observed when monochromatic radiation is incident upon optically transparent (negligible absorption) media. In addition to the transmitted light, a portion of the radiation is scattered. Most of the light that is scattered is elastically scattered at the same wavelength; however, some of the incident light of frequency $\omega_{o}$ exhibits inelastic scatter with frequency shifts $\pm \omega_{m}$, which is associated with transitions between rotational, vibrational and electronic levels. In general, the intensity and polarization of the scattered radiation are dependent upon the position of observation relative to the incident energy. Most studies use the Stokes type of scattering bands, which correspond to the $\omega_{o} - \omega_{m}$ scattering. Therefore, the Raman bands of interest are shifted to longer wavelengths relative to the excitation wavelength. An example of this is depicted in Figure for water-subtracted glucose. 

Compared with near-infrared spectroscopy, Raman spectroscopy monitors fundamental vibrations, which are sharper and exhibit less overlap. Moreover, water has a low Raman cross-section, although its infrared absorption capacity is high. However, scattering effects and the re-absorption of light in bio-tissues make the detection of Raman signals a difficult task. Protein molecules, for example, produce a background fluorescence signal that is often equal to or larger than the Raman signal itself. For these reasons, the anterior chamber of the eye and aqueous humour are the best sites for noninvasive Raman measurements. Unfortunately, these are sensitive parts of the body, and the signal level has to be low such that the power of incident irradiation is confined to a safe dose. By analysing the resulting spectra of Raman scattering it is possible to get information about the chemical structure (including glucose) of the medium. \cite{Lambert,Berger,Goetz}. Recent \cite{Enejder} data shows demonstrate the feasibility of Raman spectroscopy in monitoring blood glucose \textit{in vivo}.
It reports the first successful study of use of Raman spectroscopy for quantitative, noninvasive transcutaneous measurement of blood analytes, using glucose as an example. As an initial evaluation of the ability of Raman spectroscopy to measure glucose transcutaneously. Raman spectra were collected transcutaneously along with glucose reference values provided by standard capillary blood analysis. A partial least squares calibration was created from the data from each subject and validated using leave-one-out cross validation. The mean absolute errors for each subject were 7.8\% $\pm$ 1.8\% (mean $\pm$ std) with $R^{2}$ values of 0.83$\pm$0.10. Spectral evidence provides that the glucose spectrum is an important part of the calibrations by analysis of the calibration regression vectors.

\begin{figure}[ht]
\centering{\includegraphics[width=5in]{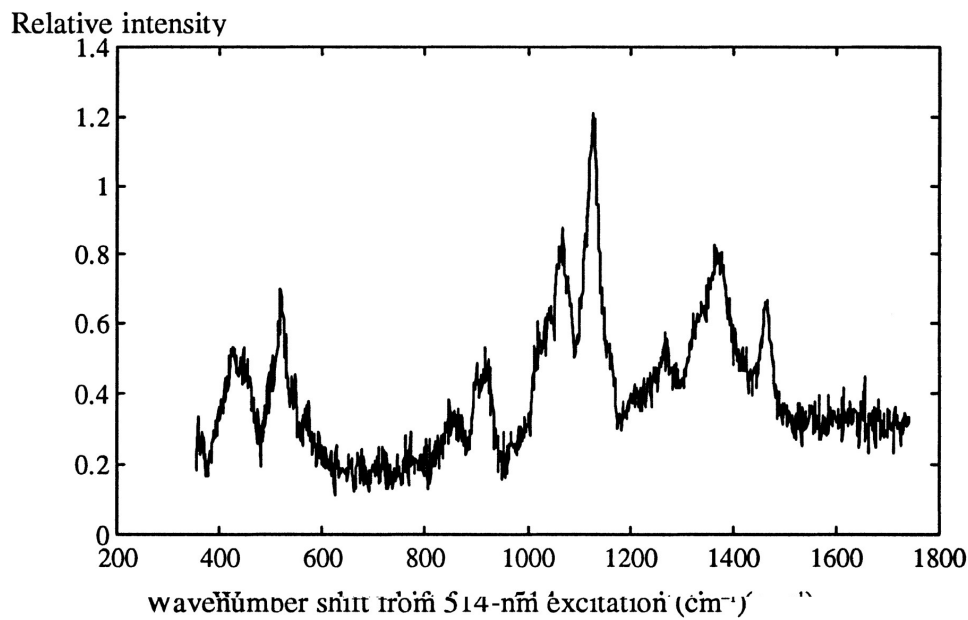}}
\caption{\sf Typical Stokes Raman spectrum is shown for glucose as a vibrational intensity versus shift in wave numbers from the 514-nm excitation wavelength. The water background has been subtracted and the background sample autofluorescence has also been removed for this nearly saturated glucose concentration.} 
\label{oct6}
\end{figure}

As with IR spectroscopic techniques, Raman spectra can be used to identify molecules because these spectra are characteristic of variations in the molecular polarizability and dipole moments. Raman spectroscopy can be considered as complementary to absorption spectroscopy because neither technique alone can resolve all of the energy states of a molecule; in fact, for certain molecules, some energy levels may not be resolved by either technique. Due to the anharmonic oscillator model for dipoles, overtone frequencies exist in addition to fundamental vibrations. It is an advantage of Raman spectroscopy that the overtones are much weaker than the fundamental tones, thus, contributing to a simpler spectra compared with absorption spectroscopy. 
One advantage to used Raman spectroscopy in biological investigations is that the Raman spectrum of water is weak, which, unlike IR spectroscopy, only minimally interferes with the spectrum of the solute and thus, the spectrum can be obtained from aqueous solutions. Replacement of slower photomultiplier tubes with faster charged coupled device arrays as well as the  higher power NIR laser diodes, makes it now possible to distinguish tissue types and quantifying blood chemicals in real time. In some studies eye has been suggested as a site for glucose concentration measurements. The reason, being the reduction the high fluorescence background, which is incurred in heavily vascularized tissue, due to the high concentration of proteins and other fluorescent components. Investigators have applied statistical methods such as partial least squares for estimation of biochemical concentrations from Raman spectra. These statistical methods combined with more affordable instrumentation give Raman the potential to also be a viable noninvasive glucose sensor. 

There are disadvantages to used the eye for Raman spectroscopy studies, with the primary concern being the laser excitation powers. The power must be kept low to prevent injury, but this significantly reduces the signal-to-noise ratio. In addition, in other tissues, especially those with blood, a large background fluorescence overwhelms the Raman signal. Instrumentation to excite in the NIR wavelength range has also been proposed to overcome this problem because the fluorescence component falls off with increasing wavelength. Excitation in the NIR region also offers longer wavelengths, which pass through larger tissue samples with lower absorption and scatter than other spectral regions such as visible or ultraviolet. However, in addition to fluorescence falling off with wavelength, the Raman signal also falls off to the fourth power as wavelength increases. Thus, there is a trade off between minimizing fluorescence and maintaining the Raman signal. In addition, like IR and NIR absorption, to quantifiably determine the inherently low concentrations of glucose \textit{in vivo} one also must account for the presence of different chemicals that yield overlapping Raman signals.

\subsection{Photoacoustic and optoacoustic techniques }
The photoacoustic (PA) technique is based on the detection of pressure waves generated by absorbing photons. A conventional method for studying gases and liquids is to generate pressure waves with a continuous wave light source and a chopper \cite{Tam}. Another possibility is to use pulsed light, such as pulsed lasers, as an energy source. Nanosecond (ns) range optical pulses are used to induce a rise in temperature, and thus a volumetric expansion inside the studied sample. The pressure waves generated this way can be detected with acoustic or optical detectors. The optical detectors are based, for example, on a probe-beam deflection method or on an interferometer \cite{Christison}. In the probe beam deflection method the probing light beam deflects when traveling through a region of refractive index change affected by the PA pulses. The interferometer, on the other hand, is very sensitive in detecting PA pulse-induced pressure changes on the surface of the sample. The detection of glucose with the PA technique is based on registering the changes in the peak-to-peak value of the signal, whereas the optoacoustic (OA) technique is based on analysing changes in the exponential curve fitted to the time-domain pulse profile. The PA and OA techniques have been used to measure the glucose content of the human body and blood vessels \cite{Quan}.

\begin{figure}[ht]
\centering{\includegraphics[width=5in]{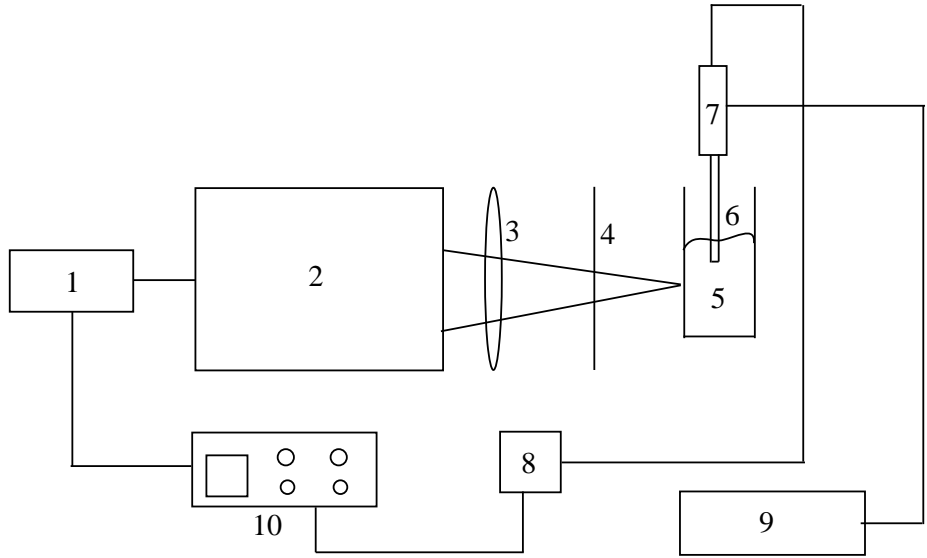}}
\caption{\sf Photoacoustic measurement system. 1: Laser unit, 2: Laser resonator, 3: Collimating lens, 4: Filter, 5: Cuvette, 6: Acoustic transducer, 7: Photoacoustic
preamplifier, 8: Main amplifier, 9: Power unit, and 10: Oscilloscope.} 
\label{oct7}
\end{figure}

Compared with optical absorption techniques, the photoacoustic methods offer the advantage of higher detection sensitivity, as the PA signal is influenced not only by the optical absorption coefficient, but also by other physical parameters including thermal expansion, specific heat and acoustic velocity. This has been demonstrated in the determination of traces and oils in liquids. The basic principle behind the PA mechanism is that an energy source (a pulsed laser, for example) irradiates some absorbing material, causing a fast thermal expansion in the illuminated volume. The energy of the expansion is released as an acoustic wave. A research group led by Mackenzie applied a pulsed PA method to measure blood glucose. They employed near-infrared PA spectroscopy to the study of glucose aqueous solutions and blood components. To demonstrate the possibility of non-invasive detection of glucose in the human body, Quan carried out an experiment using a gelatine-based tissue phantom and a circulation system containing a glucose solution. The reported detection sensitivity at 1.064 $\mu m$ was 0.071\% per mmol/l. At the same time, Christison \cite{Hayward} applied a hybrid pulsed TEA CO$_{2}$ laser to the detection of physiological glucose concentrations (18-450 mg/dl) in human whole blood. The achieved sensitivity was quite good. A portable non-invasive blood glucose monitor based on the PA method was developed with laser diodes emitting at the 904 nm wavelength. Having tested the apparatus, the group reported a correlation coefficient of 0.967 in observed blood glucose concentrations between the PA response and hospital tests on a venous blood sample \cite{Bednov}. In Germany, a research team headed by Spanner also used PA and optical techniques to investigate hemoglobin and glucose in the human body \cite{Bednov}. It is worth noting that they used a special modulating array of laser diodes. Recently, the correlation between glucose concentration and the reduced scattering coefficient of tissue has been put forward and confirmed. A research group lead by Oraevsky established that 1 mmol/l increase in glucose concentration resulted in a 3\% decrease in the optical attenuation of a rabbit's sclera \textit{in vivo}. The result was obtained by measuring a laser-induced acoustic profile using the time-resolved PA technique at 355 nm \cite{Bednov}. This sensitivity far exceeds any values achieved in previous \textit{in vivo} glucose measurements.
\subsection{Spatially resolved diffuse reflectance measurements }
The spatially resolved diffuse reflectance technique uses a narrow beam of light to illuminate a restricted area on the surface of section under study. Diffuse reflectance is measured at several distances from the illuminated area. The intensity of this reflectance depends on both the scattering and absorption coefficients of the tissue. Reflectance measured in the immediate vicinity of the illuminated point is mainly influenced by scattering of the skin, while reflectance farther away from the light source is affected by both scattering and the absorption properties of skin. The recorded light intensity profiles are used to calculate the absorption coefficient ($\mu_{a}$ and the reduced scattering coefficient ($\mu_{s}'$) of the tissue based on the diffusion theory of light propagation in tissue \cite{Heinemann}. Because $\mu_{s}'$ and glucose concentration are correlated, the latter can be extracted by observing changes in the former \cite{Heinemann}. Brulsema et al \cite{Bruulsema}carried out a glucose clamp experiment to measure the diffuse reflectance. An optical probe was fixed on the patient's abdomen. The clamping protocol consisted of a series of step changes in blood glucose concentration from the normal level of 5 mM to 15 mM and back to 5 mM. Three different clamping experiments were done on the body of diabetic volunteers at wavelength of 650 nm. The corresponding changes
in $\mu_{s}'$ were estimated to be about -0.20 percent/mM, -0.34percent/mM and -0.11 \% /mM, respectively. 
A qualitative correlation between the estimated change in $\mu_{s}'$ and the change in blood glucose concentration was observed in 30 out of 41 diabetic volunteers. Heinemann \cite{Bruulsema} obtained a similar result (-1.0 \% / 5.5 mM) in their glucose clamp experiments. However, under normal conditions, blood glucose concentration does not change as rapidly as the clamping experiments suggest. Hence, OGTT was applied in the measurement of diffuse reflectance in reference \cite{Bruulsema} at the wavelength of 800 nm. The results indicated a mean relative change in $\mu_{s}'$ of about -0.5 \%/mM and 0.3\%/mM for healthy persons and type II diabetic patients, respectively. The acceptable correlation between blood glucose concentration and $\mu_{s}'$ was 75\% (27 out of 36 measurements).
\begin{figure}[ht]
\centering{\includegraphics[width=5in]{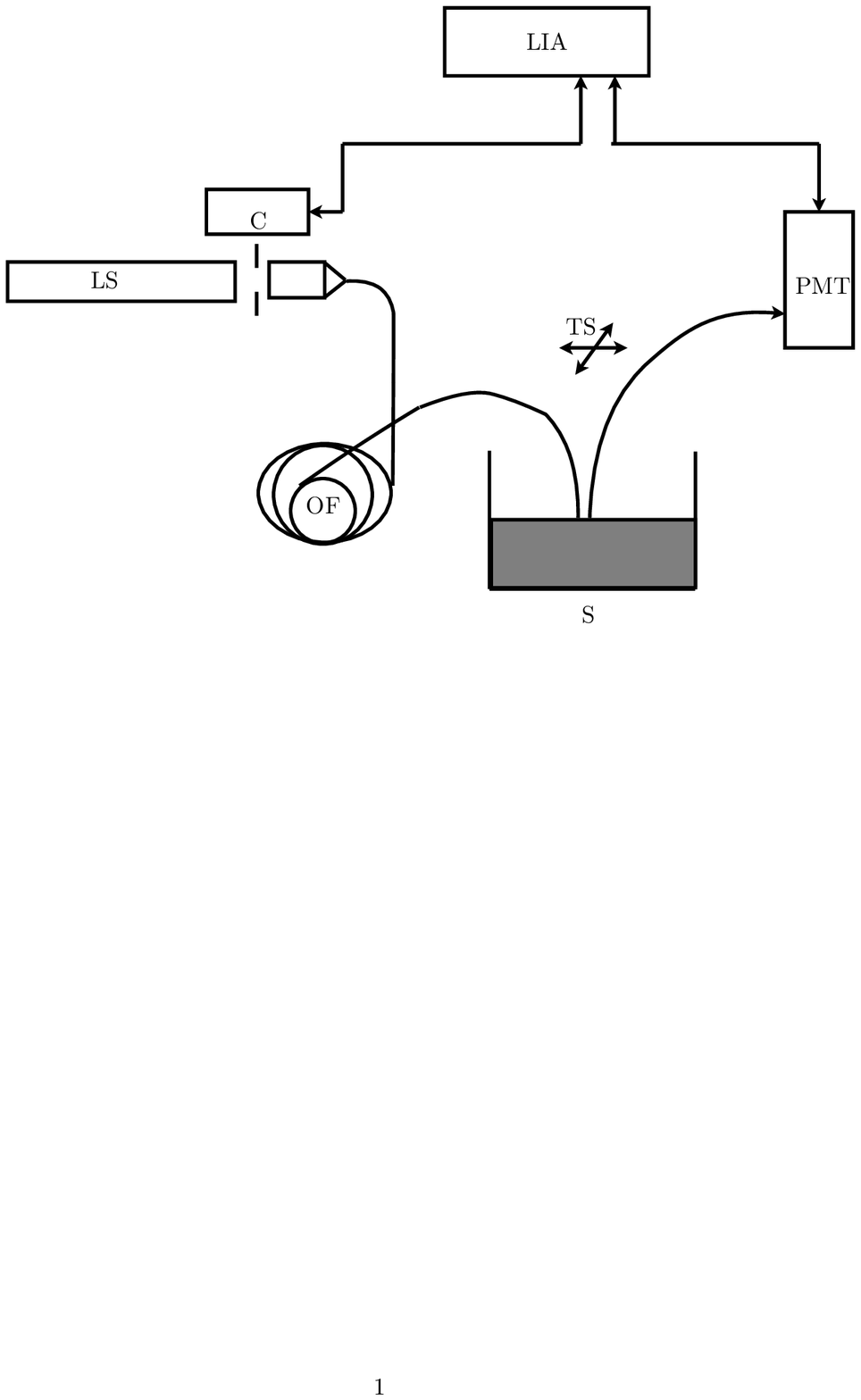}}
\caption{\sf Schematic of the experimental setup used (Poddar et al) for collection of backscattered signal. LS - light source, C - Chopper, OF - Optical fiber, TS - stepper motor controlled translation stage, LIA - Lock-in amplifier, S - sample under study and PMT - Photomultiplier tube.} 
\label{oct8}
\end{figure}
The recent results by Khalil et al. \cite{Khalil} measured with the spatially resolved diffuse reflectance technique show that temperature affects the cutaneous scattering coefficient $\mu_s'$ and absorption coefficient ($\mu_a$) values. Cutaneous $\mu_s'$ shows linear changes as a function of temperature whereas the changes in ($\mu_a$) showed complex and irreversible behavior. The thermal response of skin has been used as a basis for non-invasive differentiation of normal and diabetic skin \cite{Yeh}.

 Recently Poddar et al \cite{Poddar} shown a logarithmic correlation between BGC and $\mu^{'}_{s}$. They interpret data in terms of Monte-Carlo simulation to find values of $\mu^{'}_{s}$, $\mu_{a}$ and $g$.
\subsection{Frequency-domain reflectance technique }
The optical system used in frequency-domain reflectance measurements is similar to that in above diffuse reflectance measurements, except that the light source and the detector are modulated at a high frequency \cite{Patterson}. Then, the phase and intensity of the photon-density wave generated by the source are measured. Combining these measurements with linear transport theory enables the deduction of $\mu_a/n$ and $n \mu_s'$, where $n$ is the mean refractive index of the tissue. Maier et al \cite{Maier} applied the frequency-domain tissue spectrometer to do an OGTT for a non-diabetic male. The optical source had a wavelength of 850 nm and the measurement location was muscle tissue in the subject's thigh. The experimental results showed that the relative change of $\mu_s'$ with blood glucose concentration was - 2.5 \% / 3.6mM, which is identical with the results obtained by Heinemann.
\subsection{Polarization changes}
The basis of this optical approach is that the linear polarization vector of light will rotate when the light is passed through a substance and that the rotation measured is proportional to the concentration of the substance being monitored. This rotation is due to a difference in the indice of refraction $n_{L}$ and $n_{R}$ for left and right circularly polarized light passing through a solution containing the molecule. It occurs by virtue of the molecule's chirality or asymmetry, by which we mean the molecule has at least one center about which its mirror image cannot be superimposed upon itself. A variety of both polarimeters, adopted to the examination of all chiral substances and saccharimeters designed solely for polarizing sugars, have been developed. Glucose in the body is dextrorotatory (rotates light in the right-handed direction with concentration). In addition to the concentration of the chiral material, the amount of rotation of linear vector of the polarized light also depends on (i) the thickness of the layer traversed by the light, (ii) the wavelength of the light used for the measurement,(iii) the temperature and (iv) the pH of the solvent. Historically, polarimetric measurements have been generally obtained under a set of standard conditions. The path length typically used as a standard in polarimetry is 10 cm for liquids, the wavelength is usually that of the green mercury line (5461 Angstroms) and the temperature is 20 $^{\circ}$C. If the layer thickness in decimeters (0.1 m) is $L$, the concentration of solute in grams per 0.1 L of solution is C,  is the observed rotation in degrees, and  is the specific rotation or rotation under standard conditions, which is unique for all chiral molecules, then 
\begin{equation}
C = 100[\alpha]/L \alpha]
\end{equation}
In the above equation the specific rotation ($[\alpha]$) of a molecule is dependent upon temperature, wavelength and the pH of the solvent. Of these three, the wavelength of the light has the dominant effect on the specific rotation. This fact could potentially be used to distinguish chiral rotation of the molecule of interest from other confounding molecules as well as chiral rotation from birefringence due to the tissue.

For polarimetry to be used as a noninvasive technique for any chiral molecule and in particular for blood glucose monitoring, the signal must be able to pass from the source, through the body and to a detector without total depolarization of the beam. Because the skin possesses high scattering coefficients, maintaining polarization information in a beam passing through a thick piece of tissue (i.e., 1 cm), which includes skin, would not be feasible. Tissue thicknesses of $\leq$ 4 mm, which include skin, may potentially be used, but the polarimetric sensing device must be able to measure millidegree rotations in the presence of $\geq$ 95 \% depolarization of the light due to scattering from the tissue. As an alternative to transmitting light through the skin, several investigators have suggested the eye, as depicted in Figure , as a site for detection of \textit{in vivo} glucose concentrations. For instance, an observed rotation of 4.562 millidegrees per optical pass can be expected for a normal blood glucose level of 5.55 mmol/L, given a specific rotation of glucose at a wavelength of 633 nm of 45.62$^{\circ}$dmg/ml and thickness of 1 cm. A path length on the order of 1 cm is considered, because this is the approximate width of the average anterior chamber of a human eye.

The first optical glucose-sensing approach using polarization rotation of light through the eye was described by Rabinovitch and co-workers \cite{Rabinovitch}. The approach used a single wavelength amplitude-based technique. In this work, it was found that the glucose concentration in the aqueous humor was two orders of magnitude higher than any other optically active substances for the rhesus monkey animal model. In addition, limited data were taken to show that the measurement of glucose concentration of the aqueous humor of the eye correlated well with blood glucose levels, with a minor time delay (on the order of minutes), in rabbit models. Coté et al. \cite{Coté} developed an open-loop phase technique to increase the signal-to-noise ratio of the sensor and theoretically account for potential noise sources anticipated in the \textit{in vivo} system. King et al. (1994 )\cite{King}developed a closed-loop system using a Pockels cell, which, when used with a multi-wavelength light source, could potentially compensate for birefringence of the polarized light due to the cornea and interference due to other optically active components. Cameron and co-workers \cite{Cameron} developed a digital feedback approach, which increased the robustness and repeatability of previous polarimetric systems and demonstrated measurement of glucose in aqueous cell culture media. Chou et al.\cite{Chou} implied that the time lag between blood and aqueous humor was on the order of 30 min. 
\begin{figure}[ht]
\centering{\includegraphics[width=3in]{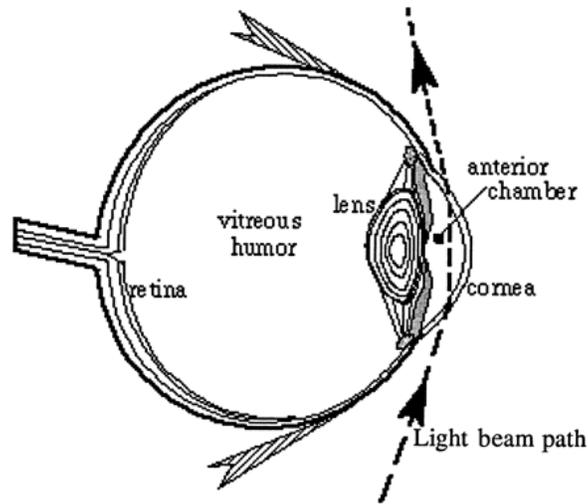}}
\caption{\sf Schematic diagram of polarization sensitive change in human eye.} 
\label{oct9}
\end{figure}
The key technological problem to be overcome before this approach is viable for glucose monitoring in the eye is the confounding rotation due to corneal birefringence and the variation in this rotation with eye motion artifact. As shown in equation 4, the rotation is directly proportional to the path length and, thus, it is critical that this length be determined or at least kept constant for each individual subject regardless of the sensing site. This can potentially be overcome in two ways. The first is to use multiple polarization states (linear polarization at $\pm$~ 45 $^{\circ}$ and left and right circular polarization) to separate birefringence from chiral rotation as described by the full Jones or Mueller matrix theory. Finally, there is the possibility that multiple wavelengths could be used because the rotation due to glucose will vary with wavelength differently than the birefringence. 

Very recently, a highly sensitive optical polarimetric
sensing system has been demonstrated\cite{apc} using method
of Orthogonal Twin Polarized Light (OTPL), which converts
micro-angle rotated by optical active substance such as glucose
to energy difference of OTPL By detecting the small polarization
rotation of polarized laser light passing through a glucose containing
fluid, this sensing system is suitable for noninvasive
glucose monitoring for diabetes patients. A resolution glucose
level of 40 mg/dl has been obtained, with a 0.9777 correlation
coefficient between the glucose concentration and the measured
values. The advantages of this system are that it can
make use of visible light, reducing influence of phase significantly,
easily available and it can be easily miniaturized.

\subsection{Fluorescence}
Several reports of fluorescence-based detection and assays have also appeared in the literature for a variety of chemicals including folate \cite{McShane}, retinal binding protein for vitamin-A status \cite{McNichols} and glucose. The fluorescence approach is different from the other optical approaches described in that it requires the sample be in contact with the sensor and, thus, cannot be developed as a totally noninvasive technology but rather requires fluid extraction or an implant. For folate and vitamin A, the fluorescence approaches have been used in the central laboratory as a process that includes high performance liquid chromatography, and only recently have technologies for monitoring the vitamin A and glucose levels been investigated for development into a field instrument.
The results reported by Pickup et al \cite{Pickup}show that intrinsic tissue fluorescence could possibly be used as a basis for non-invasive glucose monitoring. This technique is based on the detection of the fluorescent cofactor nicotinamide adenine dinucleotide phosphate (reduced) (NAD(P)H), a product of glucose metabolism.When exciting at 340 nm NAD(P)H has fluorescence at a wavelength of 440 – 480~nm. Fluorescence-based sensors are very sensitive and they can even measure glucose at the molecular level. They have been tested in \textit{in vitro} models and they need to be studied in \textit{vitro}. The sensitivity of fluorescence is necessary because this fluid has a glucose concentration range that is many orders of magnitude smaller (micromolar) than that of blood glucose. 

These fluorescent approaches generally fall into two categories: the glucose oxidase (GOD)-based sensors and the affinity-binding sensors \cite{Russell}. In the first category, the sensors use the electroenzymatic oxidation of glucose by GOX to generate an optically detectable glucose-dependent signal. Several methods for optically detecting the products of this reaction and, hence, the concentration of glucose driving the reaction have been devised. All these approaches have been explored with short-term use in mind because they all use indwelling fiber optic probes. The primary drawback to GOD based sensors is that their response depends not only on glucose concentration, but also on local oxygen tension in \textit{in vivo} conditions. Both fluorescence intensity and lifetime detection need further investigation \cite{Pickup}. 

\begin{figure}[ht]
\vskip-3cm
\centering{\includegraphics[width=4.5in]{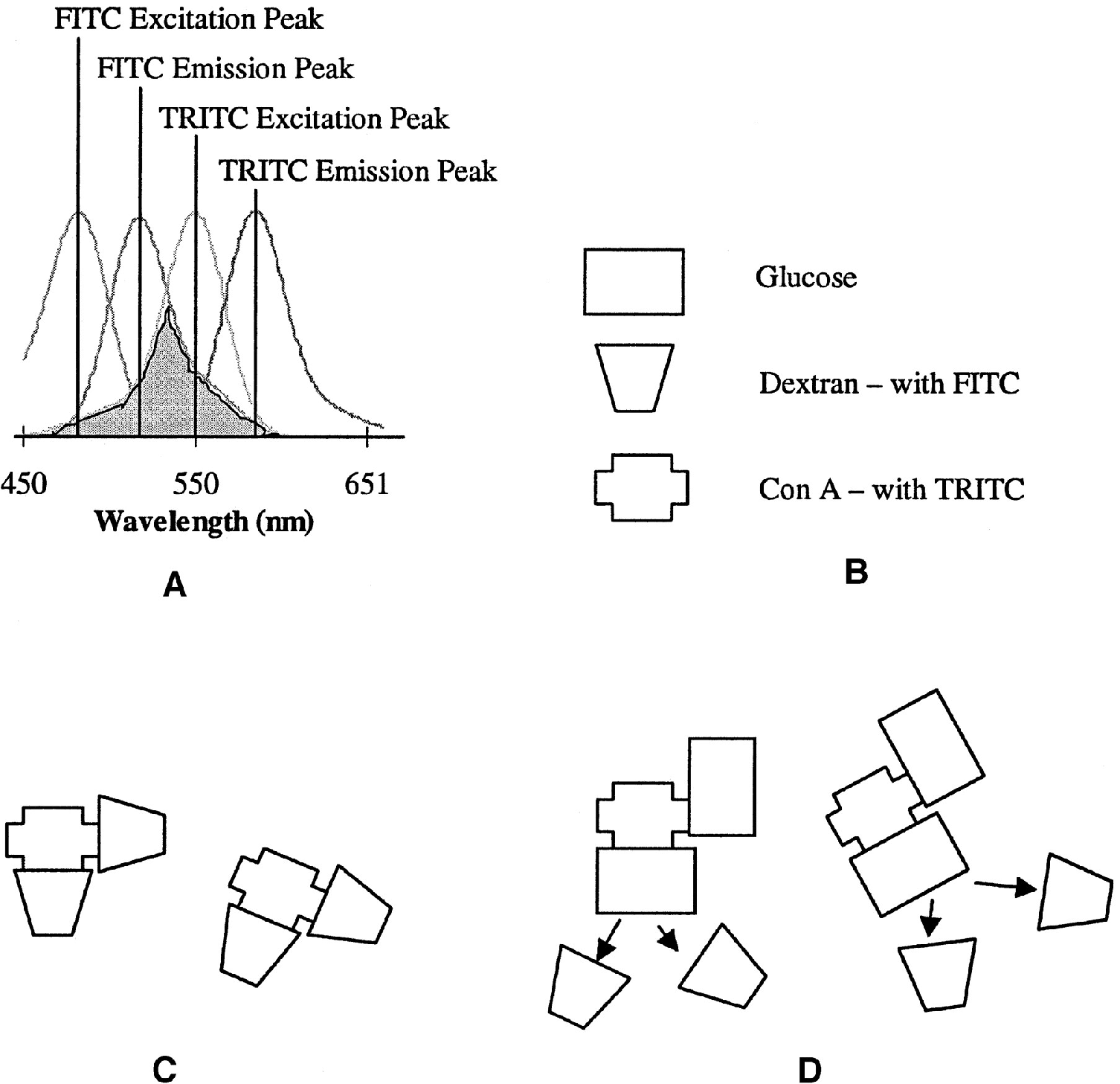}}
\caption{\sf Pictorial depiction of the phenomenon of FRET. A, The excitation and emission spectra for both the FITC and TRITC fluorephores are shown and, as depicted, the emission spectra of FITC overlaps with the excitation of TRITC. B, Cartoon characters for the glucose, TRITC-labeled Con A and FITC-labeled dextran. C, This shows the no glucose case in which the FITC-labeled dextran is bound to the TRITC-labeled Con A so that the two fluorephores are in close proximity, thus producing a quenching of the FITC emission peak. D, In the presence of glucose the FITC-labeled dextran is displaced, is no longer in close proximity to the TRITC-labeled Con A, and, therefore, the emission peak of FITC would rise.} 
\end{figure}

The affinity-based sensors do not depend on local oxygen; however, many of the earlier affinity-binding techniques were investigated for short-term use because they required indwelling probes. The most prominent fluorescence approaches have exploited the concanavalin A (Con A) affinity for polysaccharides. Immobilized Con A was used as a receptor for competing species of fluorescein isothiocyanate (FITC)-labeled dextran and glucose. Increased concentrations of glucose displace FITC-dextran from Con A sites, thus, increasing the concentration and fluorescence intensity of FITC-dextran in the visible field. The phenomenon of fluorescence resonance energy transfer (FRET), whereby an acceptor in close proximity to a fluorescent donor can induce fluorescence quenching in the latter as shown in Figure  . In most of the reported literature, glucose detection based upon FRET was between FITC-bound dextran and tetramethylrhodamine isothiocyanate (TRITC)-bound Con A. When TRITC-Con A is added to a solution of FITC-dextran, the binding of the dextran to the Con A results in the required molecular proximity $(54 A^{\circ})$ for FRET-based quenching to occur. Mansouri and Schultz (1984 )\cite{Schultz} reported that glucose concentrations could be measured in aqueous solutions by a proportional change in FITC fluorescence. The technique was both very specific to glucose and sensitive to glucose concentration, without interference from other constituents frequently found in blood plasma.

Lakowicz and co-workers \cite{Lakowicz} developed similar fluorescent assays for glucose, based on phase-modulation fluorimetry and Con A-dextran moieties. The authors used fluorescence lifetime techniques and FRET to indirectly measure glucose concentrations. They have more recently devised a similar sensor in which ruthenium-Con A and maltose-insulin-malachite green are used as the reagents. Increased glucose concentration causes an increase in both fluorescence intensity and fluorescence lifetime of the ruthenium dye.  
One such encapsulation system, consisting of alginate-poly-L-lysine spheres that encapsulated glucose-sensitive, fluorescently labeled macromolecules is also a promising technique. Similarly, constructed micro capsules have been demonstrated to be highly permeable to water and low-molecular-weight compounds. Fluorescence intensity of FITC emission from these spheres was shown to be glucose responsive, but the dextran displacement due to competitive glucose binding was not reversible within a reasonable timescale.

Most recently,  the use of poly(ethylene glycol) (PEG) particles to encapsulate the FRET assay \cite{Pickup} is reported. This polymer has been reported to have numerous properties beneficial for use \textit{in vivo} and may potentially overcome many of the drawbacks of the alginate/poly-L-lysine system. A highly water-soluble hydrogel is formed upon cross-linking. PEG-based polymers have previously been evaluated for \textit{in vivo} use as protein drug delivery devices, for postoperative adhesion prevention and for biocompatible membranes over electrochemical sensors. PEG-based coatings were reported to improve the biocompatibility of implanted glucose sensors, without being glucose mass-transfer limiting. The stability and solubility of numerous proteins are reportedly increased upon conjugation to PEG. Con A has been conjugated to monomethoxy PEG-5000 while retaining its sugar-binding abilities. It is possible to create a micro particle-based fluorescent glucose assay system potentially suitable for subcutaneous implantation and the optimization of the glucose response through control of the Con A to dextran ratio within the gel. 

Overall, the advantage of fluorescence sensors is that they can be made highly sensitive and highly specific to the analyte of interest and eliminate many of the potential interferences common with other techniques. However, this technique require the admission of exogenous chemicals to the subject under study. Additionally, long-term studies are required to assess the extent to which these chemicals may be susceptible to degradation over time via consumption, photo-bleaching or denaturation.

\subsection{Time of Flight measurements}
Time of Flight (TOF) measurements have been adopted to measure the effect of glucose on blood \textit{in vitro} at a wavelength of 906 nm. In photon migration measurements with the TOF technique, short laser pulses are injected into the sample. The photons of the pulse undergo many absorption and scattering events when traveling in the sample. The scattering processes make the photon path lengths longer. Useful data can be obtained about the optical properties of the sample ($\mu_s$ and $\mu_a$) by observing the TOF distributions of the photons and by analysing their shapes. The calculation of different pulse parameters, such as mean time-of-flight, full width at half maximum (FWHM), integral of the pulse, center-of-gravity, and moments may help in this analysis. \cite{Yoo}. 
\begin{figure}[ht]
\centering{\includegraphics[width=4in]{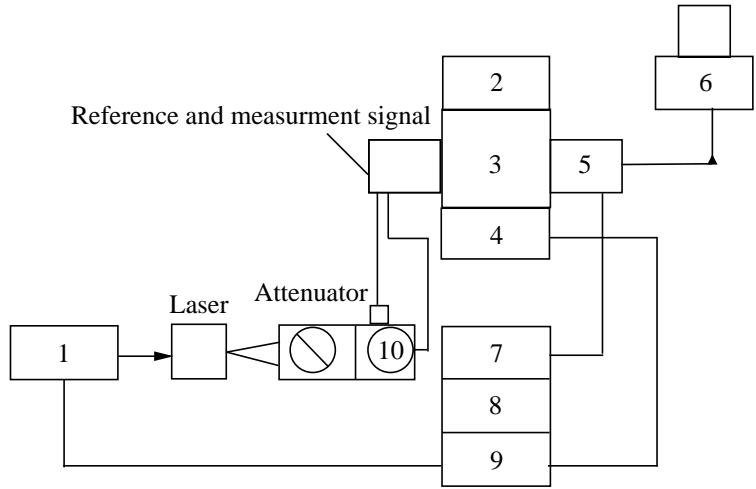}}
\caption{\sf Time-of-flight measurement system. 1: Picosecond laser module, 2: Blanking
unit, 3: Fast speed sweep unit, 4: Streak camera, 5: Digital camera, 6: PC, 7: Camera controller, 8: Power supply unit, and 9: Delay unit.} 
\label{oct11}
\end{figure}

Fig. shows an example of a TOF measurement system. A picosecond (ps) laser module with an approximately 30 ps pulse length at a wavelength of 906 nm was used as a light source and a streak camera was used as a detector in the laser pulse measurements. The energy of a pulse is 1 nJ. The detection of photons in the streak camera is done with a photocathode of the streak tube. The light of the photocathode is converted into electrons. The electrons travel through sweep electrodes, which direct them to the microchannel plate (MCP). The electrons are multiplied by the MCP and are eventually converted into light on a phosphor screen. The image on the phosphor screen, containing intensity information as a function of time, is captured with a CCD camera and shown on a computer screen. The TOF technique with a streak camera takes a long measurement time.

\subsection{Optical Coherence Tomography }

Optical Coherence Tomography (OCT) is based on the detection of back-scattered photons with an interferometer. The interferometer consists of sample and reference arms, a light source and a detector. The scanning mechanism in the reference arm enables the detection of photons from different depths in the sample. The glucose sensing is based on analysing changes in the slope value fitted to the OCT signal depth profile. Esenaliev and Larin \cite{Esenaliev} have used OCT for non-invasive glucose monitoring both \textit{in vitro}and \textit{in vivo}. OCT has better spatial and depth resolutions than the PA and TOF techniques, but a smaller imaging depth.

An OCT system with the wavelength of 1300 nm was used in 15 healthy subjects in 18 clinical experiments. Standard oral glucose tolerance tests were performed to induce changes in blood glucose concentration. Blood samples were taken from the right arm vein every 5 to 15 min. OCT images were taken every 10 – 20 s from the left forearm over a total period of 3 h. The slope of the signals was calculated at the depth of 200 – 600 $\mu m$ from the skin surface. A total of 426 blood samples and 8,437 OCT images and signals were collected and analyzed in these experiments under \textit{in vivo} conditions\cite{Larin}.

\begin{figure}[ht]
\centering{\includegraphics[width=5in]{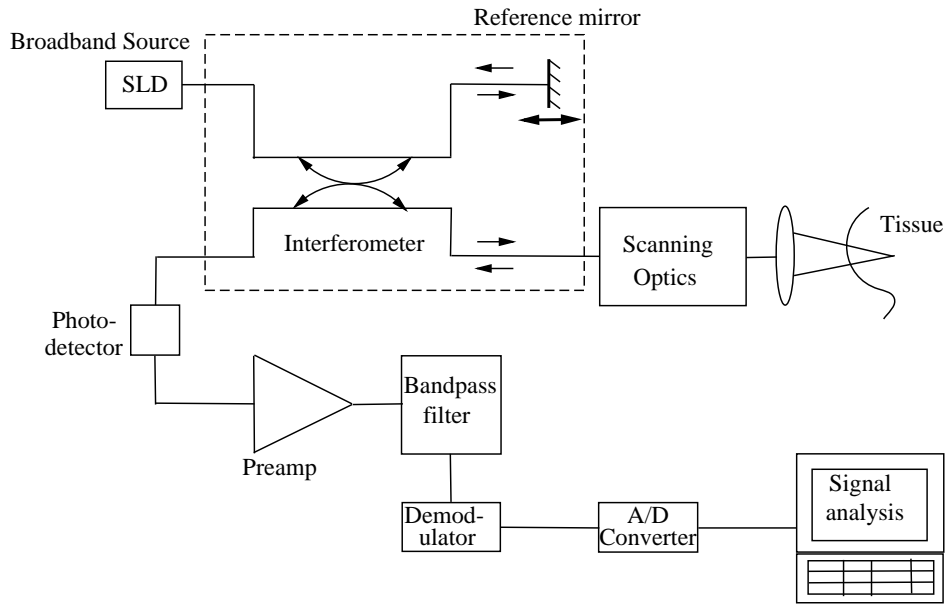}}
\caption{\sf Schematics of the experimental setup used in the clinical studies. SLD, superluminescent diode.} 
\label{oct12}
 \end{figure}  

There was a good correlation between changes in the slope of noninvasively measured OCT signals and blood glucose concentrations throughout the duration of the experiments. The slope of OCT signals change significantly (up to 2.8\% per 10 mg/dl) with variation of plasma glucose values. 

The good correlation obtained between the OCT signal slope and blood glucose concentration is due to the coherent detection of backscattered photons, which allows measurements of OCT signal from a specific tissue layer without unwanted signal from other tissue layers.
Tissue scattering properties are highly dependent on the ratio of the refractive index of scattering centers (cellmembranes, cellular components, and protein aggregates) $n_{s}$, to the refractive index of the interstitial fluid (ISF)$n_{ISF}$
Raising the tissue glucose concentration increases $n_{ISF}$ by $1.52\times10\times^{-5}$ per each 10 mg/dl, decreases the scattering coefficient ($\mu_s$) of tissues, and decreases the refractive index mismatch.

\begin{figure}[ht]
\centering{\includegraphics[width=3.5in]{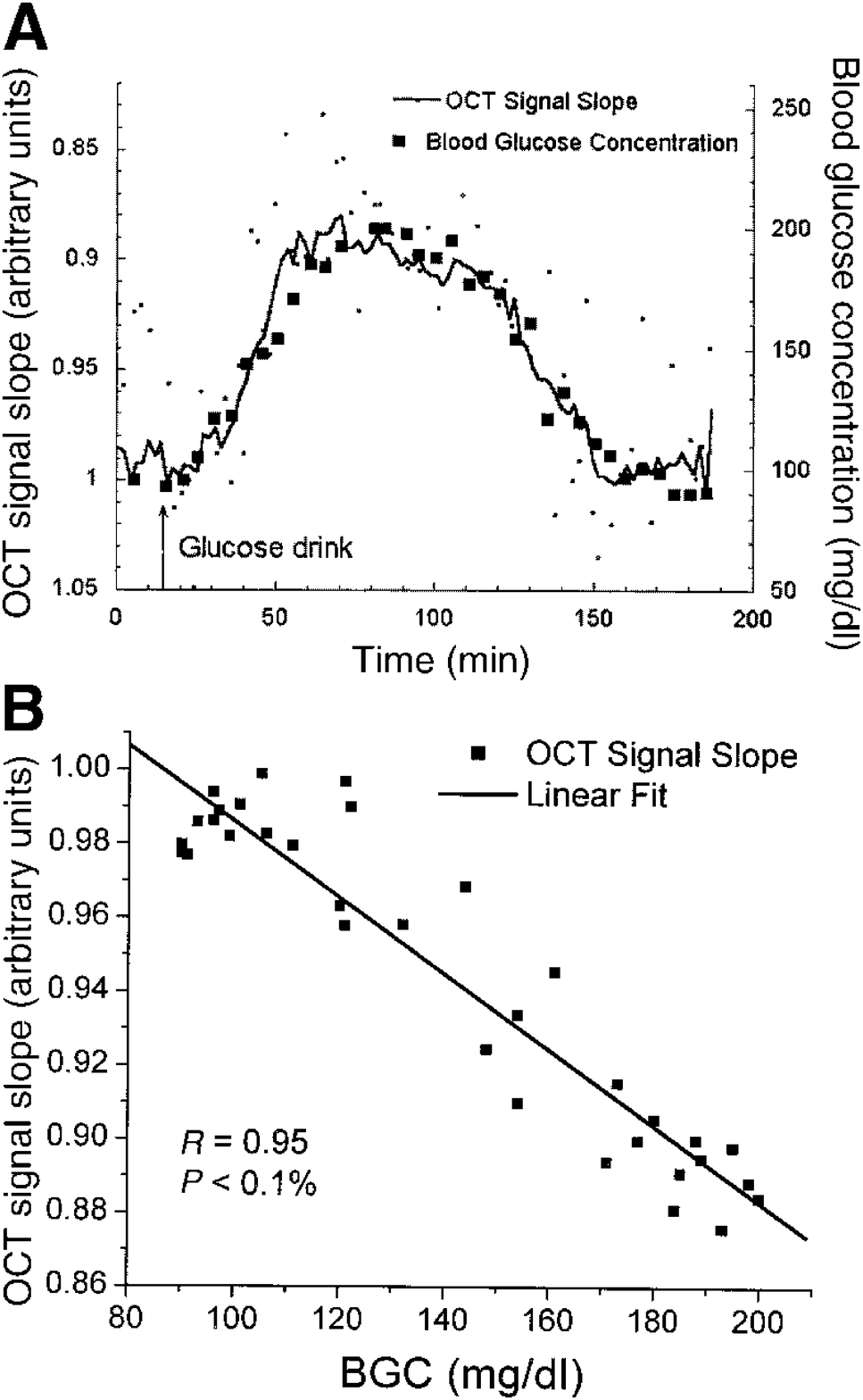}}
\caption{\sf A: Slope of OCT signals (plotted in the inverted scale) and corresponding blood glucose concentrations obtained from another healthy volunteer. The blood glucose concentration was measured every 5 min. Dots represent the OCT signal slope (in arbitrary units), and the black line represents the fit of the data points.f, actual blood glucose concentrations. B: Slope of OCT signals versus blood glucose concentration (BGC) for the data shown in A. R, correlation coefficient. Dots represent the OCT signal slope, and the line represents the linear fit of the OCT data points.} 
\label{oct13}
\end{figure}  

Recently Larin \cite{Larin1} also showed that, OCT images and signals were obtained from skin of Yucatan micropigs and New Zealand rabbits. Results obtained  demonstrate that: (i) several body osmolytes may change the refractive index mismatch between the interstitial fluid (ISF) and scattering centres in tissue, however the effect of the glucose is approximately one to two orders of magnitude higher; (ii) an increase of the ISF glucose concentration in the physiological range (3-30 mM) may decrease the scattering coefficient by 0.22\% $mM^{-1}$ due to cell volume change; (iii) stability of the OCT signal slope is dependent on tissue heterogeneity and motion artefacts; and (iv) moderate skin temperature fluctuations ($\pm 1$ $^{\circ}$C) do not decrease accuracy and specificity of the OCT-based glucose sensor, however substantial skin heating or cooling (several $^{\circ}$C) significantly change the OCT signal slope. These results suggest that the OCT technique may provide blood glucose concentration monitoring with sufficient specificity under normal physiological conditions.
We have also show\cite{Poddar1} that there is decrease in reduced scattering coefficient of blood glucose when glucose concentration is incresed \textit{in vivo} condition of human skin. We have also developed a novel mathematical model to extract reduced scattering coefficient of blood glucose. We found that the correlation obtained between the OCT signal and blood glucose concentration is due to the coherent detection of backscattered photons.

\begin{figure}[ht]
\centering{\includegraphics[width=5in]{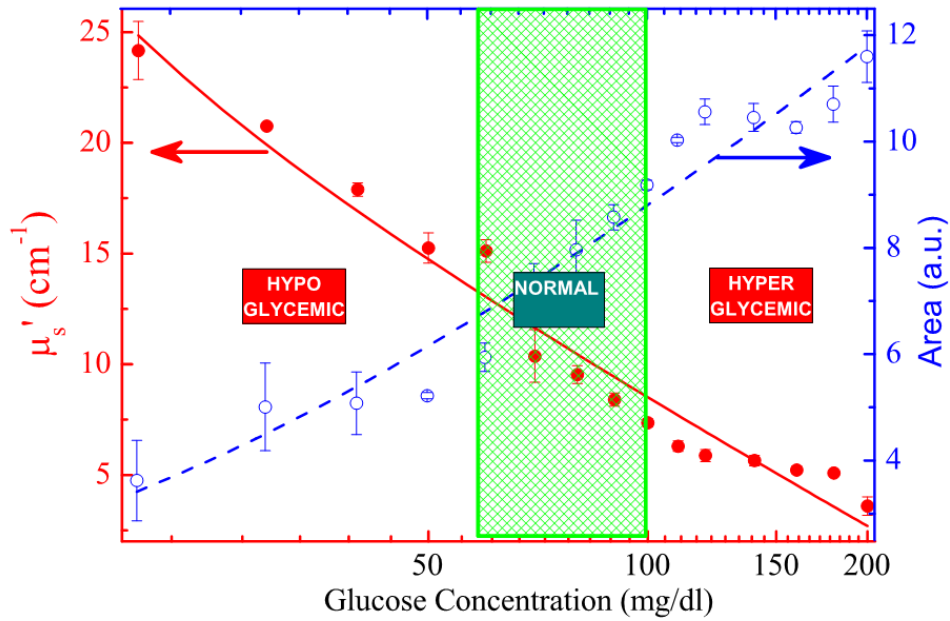}}
\caption{\sf Semilog plot of reduced scattering coefficient and curve 
area with glucose concentration. Data were extracted from Monte Carlo 
simulation and fitted to the experimental data. The error bars were obtained after 25 measurements.\cite{Poddar1} } 
\label{pod1}
 \end{figure}

Very recently, the authors have successfully demonstrated that the technique
of optical coherence tomography can be adopted for non-invassive, non-contact,
\textit{in-vivo} monitoring of blood glucose.\cite{Poddar1} Making use of changes in reduced scattering 
coefficient due to refractive-index mismatch between 
the extracellular fluid and the cellular membranes and 
armed with a theoretical model, the authors established a correlation between the glucose concentration and reduced 
scattering coefficient (Figure \ref{pod1}). The scattering coefficients are 
extracted from the deconvoluted interference signal 
using Monte Carlo simulation with valid approximations. 

\section{Fluid glucose biochemical measurement}
\subsection{Invasive methods }
\subsubsection{Biochemical and Glucometer}
Diverse enzymatic sensors based on glucose oxidation have been developed since the 1960s. The earliest blood glucose sensors were mainly used in the clinical environment or at the bedside for the continuous detection of human glucose level. These devices include implantable sensors in the form of intravenous implants, needle-like probes whose sensor
tip penetrates into the subcutaneous tissue and fully implanted subcutaneous devices requiring no communication across the skin. Shults et al. \cite{Shults1} describe a representative fully implanted sensor consisting of glucose oxidase laminated on a membrane, a sensing electrode system and a radio-telemetry transmission unit, with a total system weight of about 27 gm. Powered by a tiny lithium battery with a predicted life-span of 1.5 years, the sensor transmits signals to an external computer. Although implanted sensors may be built economically using microchip technology, they are invasive and require contact with blood. This raises concerns about their long-term stability. In addition, the head of an implanted sensor should be non-toxic and non-reactive in blood or tissue.

In the late 1960s, blood glucose measurements were introduced for h\textsf{}ome use. They enable diabetic persons to inspect their blood sugar level visual color changes on a chemical test strip or by displaying a reading on a glucose meter obtained from a drop of blood. The measuring principle is based on glucose oxidation and electrochemical methods. Usually, a few micro-liters of blood is taken by means of a finger stick or an arm stick. The blood is immediately placed on a test chip that is already inserted in a glucose meter. The meter displays the glucose concentration reading in a few tens of seconds. The main disadvantage of glucose meters is that they require patients to prick their skin, causing slight physical pain and harming the skin and the capillaries.

It should be noted, however, that a new silicon micro-needle technique has been developed by \textit{Kumetrix Inc.}\cite{Kimura} which gives a minute pain equivalent to a mosquito bite. Normally, humans are quite unaware of mosquito bites, they only feel a little itch and experience some swelling at the location of the bite, caused by enzymes that the insect uses to stop blood from coagulating. Although lacking these chemical agents, the silicon micro-needle is able to penetrate the skin and draw a blood sample painlessly, because it has a diameter that is smaller than that of a human hair. The current system consists of a hand-held, battery-powered, electronic monitor that accepts a cartridge loaded with up to 10 disposable sampling devices. The heart of the device is a silicon chip that consists of a micro-needle and a receptacle into which the blood sample is drawn. To take a measurement, the cartridge is loaded into the monitor and pressed against the skin, causing the micro-needle to penetrate the skin to draw a very small volume of blood (one-hundredth the size of a blood drop or 100 nano-liters) into the receptacle. Chemicals inside the receptacle react with the glucose to produce a color.
The more intense the color is, the higher the blood glucose level is. The monitor analyzes the color by lasers and then displays the blood glucose concentration. 

\subsection{Permeance measurements in interstitial fluid (minimally- or non-invasive method)}

Interstitial fluid (ISF) is an extracellular fluid through which chemicals (like glucose) and proteins pass on their way from the capillaries to cells. As a consequence, the glucose concentration of ISF exhibits a degree of correlation with the glucose concentration in the capillaries. Some researchers claim that the glucose concentration of ISF is practically identical to that of blood \cite{Fischer1}, whereas other investigators maintain that the equivalence is only about 75\% \cite{Sternberg}. The difference is probably attributable to the speed of the blood flow: as the glucose concentration increases, the time delay between observed changes in blood glucose and ISF glucose is found to be 12 minutes. At any rate, glucose concentration in the ISF can be used to provide an indicator of the person's blood glucose value. ISF resides just under the skin, but the low permeability of the epidermal keratinised layer blocks the permiation of the fluid through the skin. However, during the past few years only few techniques have been developed for drawing glucose from the skin.

\textit{SpectRx Inc.} in Norcross, Georgia, has also developed a smart method for accessing ISF in the skin. The method applies a low-cost, low-energy laser to create micropores in the stratum corneum. The depth of these micropores is only about 20 $\mu m$, just the thickness of stratum corneum, the layer of dead, nonviable epidermal cells forming the outer skin surface. The micropores' diameter is approximately equal to that of a human hair. The method offers several advantages. Firstly, the procedure is completely painless. Secondly, it allows an analysis of the actual ISF, not just partial compositions there of, thereby enabling a direct glucose concentration measurement by conventional glucose assay techniques such as the glucose oxidase method. The continuous monitoring system is usable for three days at a time. The test patch can be placed on almost anywhere (arms, abdomen and legs) on the human body with no loss of accuracy.

Finally, and most importantly, the procedure does not require calibration from a finger stick blood measurement. Clinical tests involving 20 diabetic patients show that the correlation between \textit{SpectRx} ISF measurements and blood glucose levels is as high as 0.90 in the 60- 400 mg/dl glucose range. Another, relatively similar method has also been reported \cite{Sternberg}. In this method, the stratum corneum is first stripped, where after effusion ISF is sampled by suction and its glucose concentration is measured by an ion sensitive FET. 

Ultrasound, especially low frequency ultrasound, can induce cavitation in and around the skin. The oscillation and collapse of cavitation bubbles disorder the lipid bilayers of the skin, resulting in enhanced skin permeability. The process is known as sonophoresis.
It was first used to enhance the transdermal delivery of drugs, but its application range was soon extended to include the transdermal monitoring of glucose and other analytes \cite{Kost}. Accomplishing a transdermal glucose flux through the skin involves two steps.
First, the skin is pre-treated by a low frequency ultrasound (20 kHz, for instance). Second, glucose is extracted using passive diffusion, low intensity ultrasound or a vacuum. Without ultrasonic pre-treatment, the permeability of glucose through a rat's skin is less than 4 $\pm$ 3 $\mu m/hr$. Ultrasonic pre-treatment and extraction increase this permeability to 0.026 $\pm$ 0.011cm/hr (or 0.034 $\pm$ 0.025 cm/hr for vacuum extraction), representing 100-fold enhancement. Clinical data shows that the skin retains a high level of permeability for about 15 hours after the ultrasonic treatment. The correlation coefficient between monitored transdermal glucose level and venous glucose level is about 0.9. The time delay between these glucose levels is about 12 minutes. Patients report no pain during the ultrasound and the technique leaves no visible effects on the skin.

Other minimally invasive techniques for monitoring glucose concentration include a painless, hollow lancet used for collecting tiny drops of ISF (\textit{Integ}), a microdialysis probe, an inserted sensor (\textit{MiniMed Technologie}s) and a fluorescent sensor (Sensor For Medicine). Nevertheless, all these techniques harm the skin, albeit slightly.
\subsubsection{Photonic Crystal Glucose-Sensing Material}
Alexeev et al \cite{Vladimir} recently reported a photonic crystal glucose-sensing material, which consists of a crystalline colloidal array embedded within a polymer network of a polyacrylamide-poly (ethylene glycol) hydrogel with pendent phenylboronic acid groups. They used new boronic acid derivatives such as 4-amino-3-fluorophenylboronic acid and 4-carboxy-3-fluoropheny lboronic acid as the molecular recognition elements to achieve sensing at physiologic pH values. 

The improved photonic glucose-sensing material sensed glucose in the range of the 100 $\mu mol/litre$ concentrations found in tear fluid. The detection limits were 1 $\mu mol/litre$ in synthetic tear fluid. The visually evident diffraction color shifted across the entire visible spectral region from red to blue over the physiologically relevant tear-fluid glucose concentrations. This sensing material is selective for glucose over galactose, mannose, and fructose. These glucose sensors have properties appropriate for use in such glucose-sensing applications as ocular inserts or diagnostic contact lenses for patients with diabetes mellitus. 

\subsubsection{Measurement of Glucose by Metabolic Heat Conformation Method}
Kyung Cho et. al \cite{Kyung} developed a method, named as the metabolic heat conformation (MHC) method, for the noninvasive measurement of blood glucose. The MHC method involves the measurement of physiologic indices related to metabolic heat generation and local oxygen supply, which correspond to the glucose concentration in the local blood supply. 
\begin{figure}[ht]
\centering{\includegraphics[width=5in]{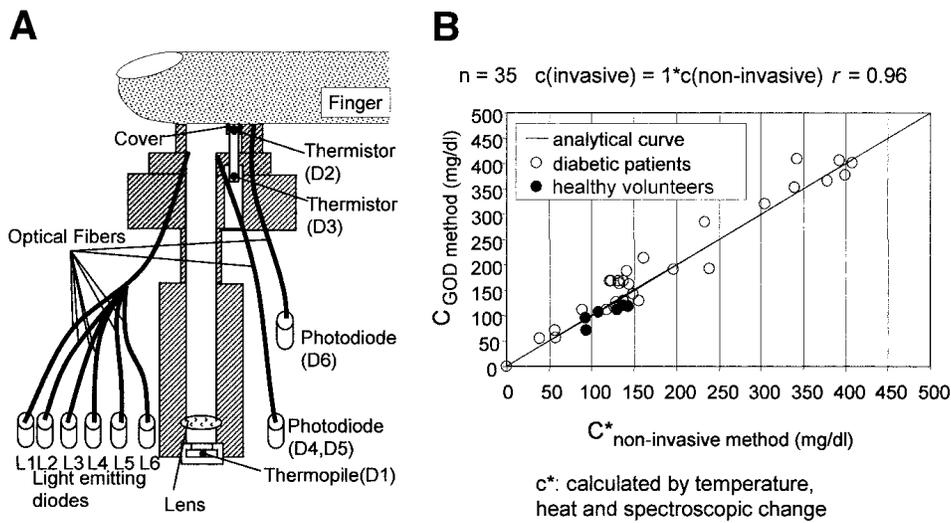}}
\caption{\sf The experimental setup. B: Regression analysis involving 35 data points (29 from
diabetic patients and 6 from healthy volunteers) by the noninvasive method against a glucose oxidase (GOD) enzymatic amperometry for whole-blood samples as the reference method. n = number of measurements. c, blood glucose concentration; r, coefficient of correlation.} 
\label{oct14}
\end{figure} 

They used noninvasive thermal and optical sensors on the fingertip of an individual to measure thermal generation, blood flow rate, hemoglobin (Hb) concentration, and oxyhemoglobin concentration. The calibration model incorporates mathematical procedures to convert signals from the sensor pickup to final glucose concentrations. The mathematical procedures are multivariate statistical analysis, involving values from sensor signals, polynomials from various values, regression analysis of individual patients, and cluster analysis of patient groups. The glucose value is calculated for each patient measurement, applying one of the clusters by discriminant analysis. Regression analysis was performed to compare the noninvasive method with the hexokinase method, using 127 data points (109 data points from diabetic patients, 18 data points from nondiabetic patients) with glucose concentrations ranging from 3.0 to 22.5 mmol/L (54 – 405 mg/dL). The correlation coefficient ($R$) was 0.91. Reproducibility was measured for healthy fasting persons; the CV was 6 percent at 5.56 mmol/L (100 mg/dL). The MHC method can be used to estimate blood glucose concentrations noninvasively.

\subsubsection{Reverse iontophoresis used in Glucowatch}
Iontophoresis is a process in which a weak electric current is employed to transport charged molecules through intact skin. The method also enables an increased transport of polar, as yet uncharged, species. The moving, charged molecules carry glucose and other chemical molecules found in the ISF through intact skin (reverse iontophoresis). Once the fluid is extracted from the skin, it can be collected and its glucose concentration is determined by glucose oxidation methods. Hence, the measurement can be performed continuously without puncturing the skin. However, the slow transit time of the fluid through the skin produces a 20-minute lag in the blood glucose reading. In addition, the glucose concentration in the fluid is much lower than in the blood. While the size of the glucose sample varies with time and the current applied, the concentration is generally in the micromolar range, while that of blood glucose is in the millimolar range. Another drawback is that the test requires periodic calibration with blood testing. As a result,
reverse iontophoresis is not a truly non-invasive method. Moreover, the accuracy of the measurement can be dramatically affected by the presence of sweat, and the low electric current passing through the skin may cause irritation.
The most successful iontophoresis-based glucose monitor is the GlucoWatch [\textit{Biograph}e, a wristwatch-type device developed by Cygnus Inc.\cite{cygn} in Redwood City, California]. The device collects glucose molecules in gel collection discs that are part of a single-use AutoSensor. containing the enzyme glucose oxidase. As glucose enters the discs, it reacts with the glucose oxidase in the gel to form hydrogen peroxide. A biosensor in contact with each gel collection disc detects the hydrogen peroxide, and generates an electronic signal. The monitor uses a calibration value entered previously by the patient to convert the signal into a glucose measurement, which is displayed on the monitor and then stored in memory. The sensor is designed to take three measurements every hour and must be replaced every 12 hours. The patient is required to perform a routing blood-based calibration every day. To avoid the sweat effect, the GlucoWatch monitor measures skin conductance, which increases with sweat. If the skin conductance exceeds a predetermined threshold, the measurement is skipped. 
\section{CONCLUSION}
Diabetes mellitus is a complex group of syndromes that have in common a disturbance in the body's use of glucose, resulting in an elevated blood sugar. Once detected, sugar diabetes can be controlled by an appropriate regimen that should include diet therapy, a weight reduction program for those persons who are overweight, a program of exercise and insulin injections or oral drugs to lower blood glucose. Blood glucose monitoring by the patient and the physician is an important aspect in the control of the devastating complications (heart disease, blindness, kidney failure or amputations) due to the disease. 
Intensive therapy and frequent glucose testing has numerous benefits. With ever improving advances in diagnostic technology, the race for the next generation of bloodless, painless, accurate glucose instruments has begun. 
However, many hurdles remain before these products reach the commercial marketplace. 

Calibration of the instruments and validation of the results obtained by the optical methods under different environmental conditions and used by different patient populations (i.e., different ages, sizes and ethnic origins) must be performed. The devices may have to be calibrated to individual users. 

Recent instrumentation lacks specificity due to substantial chemical and physical interferences. The devices use multivariate regression analyses that convert the optical signal to a glucose concentration. Large amounts of data are used to build the glucose model and must take into consideration the concentration range, sampling environment and other factors involved in the analysis. First an instrument must be designed that accurately detects glucose concentration. Correlation and clinical interpretation of this value, in respect to the patient's "true glucose" value, is imperative for optimum therapy and disease management. 

Considerable progress has been made in the development of non-invasive glucose devices however, at this time, frequent testing using invasive blood glucose determination via fingerstick provides the best information for diabetes disease management. Industry spokespersons have said: "… anyone who can come up with a viable noninvasive or painless technique is going to make a lot of money. People's lives are involved ... and we don't want to suggest that this technology is right around the corner. This is very tricky, difficult work."

\begin{acknowledgments}The authors thank DST  and AICTE for financial Support. 
\end{acknowledgments}

\end{document}